\documentclass[sigconf,anonymous=false]{acmart}

\usepackage{algorithmic}
\usepackage{graphicx}
\usepackage{textcomp}
\usepackage{xcolor}
\usepackage{makecell}

\usepackage{multirow}
\usepackage{siunitx}
\usepackage{pifont}
\usepackage{bm}

\usepackage[ruled,%
            linesnumbered]{algorithm2e}
\SetKwInput{KwInput}{Input}

\AtBeginDocument{%
  }

\setcopyright{acmlicensed}
\copyrightyear{20xx}
\acmYear{20xx}
\acmDOI{XXXXXXX.XXXXXXX}

\acmConference[Conference acronym 'XX]{Make sure to enter the correct
  conference title from your rights confirmation email}{xx--xx,
  20xx}{Woodstock, NY}
\acmISBN{978-1-4503-XXXX-X/18/06}

\usepackage{fancyhdr}   

\fancypagestyle{firstpage}{     
  \fancyhf{}
  \chead{\textcolor{gray}{This article has been accepted for publication in the ACM international joint conference on \\ Pervasive and Ubiquitous Computing (UbiComp 2024).}}
  \fancyfoot[C]{\small{\textcolor{gray}{~\copyright~ Personal use of this material is permitted.  Permission from ACM must be obtained for all other uses, in any current or future media, including reprinting/republishing this material for advertising or promotional purposes, creating new collective works, for resale or redistribution to servers or lists, or reuse of any copyrighted component of this work in other works.}}}

}

\begin{document}

\title{Earable and Wrist-worn Setup for Accurate Step Counting Utilizing Body-Area Electrostatic Sensing}

\author{Sizhen Bian}
\orcid{0000-0001-6760-5539}
\affiliation{%
  \institution{PBL, D-ITET, ETH Zürich}
  \city{Zürich}
  \country{Switzerland}
}
\email{sizhen.bian@pbl.ee.ethz.ch}

\author{Rakita Strahinja}
\affiliation{%
  \institution{PBL, D-ITET, ETH Zürich}
  \city{Zürich}
  \country{Switzerland}
  }
\email{rakitas@ee.ethz.ch}

\author{Philipp Schilk}
\affiliation{%
  \institution{PBL, D-ITET, ETH Zürich}
  \city{Zürich}
  \country{Switzerland}
  }
\email{schilkp@ee.ethz.ch}

\author{Clénin Marc-André}
\affiliation{%
  \institution{PBL, D-ITET, ETH Zürich}
  \city{Zürich}
  \country{Switzerland}
}
\email{mclenin@ee.ethz.ch}

\author{Silvano Cortesi}
\affiliation{%
  \institution{PBL, D-ITET, ETH Zürich}
  \city{Zürich}
  \country{Switzerland}
}
\email{silvano.cortesi@pbl.ee.ethz.ch}

\author{Elio Reinschmidt}
\affiliation{%
  \institution{PBL, D-ITET, ETH Zürich}
  \city{Zürich}
  \country{Switzerland}
}
\email{elio.reinschmidt@pbl.ee.ethz.ch}

\author{Kanika Dheman}
\affiliation{%
  \institution{PBL, D-ITET, ETH Zürich}
  \city{Zürich}
  \country{Switzerland}
}
\email{kanika.dheman@pbl.ee.ethz.ch}

\author{Michele Magno}
\affiliation{%
  \institution{PBL, D-ITET, ETH Zürich}
  \city{Zürich}
  \country{Switzerland}
}
\email{michele.magno@pbl.ee.ethz.ch}

\renewcommand{\shortauthors}{Trovato et al.}

\begin{abstract}

Step-counting has been widely implemented in wrist-worn devices and is accepted by end users as a quantitative indicator of everyday exercise. However, existing counting approach (mostly on wrist-worn setup) lacks robustness and thus introduces inaccuracy issues in certain scenarios like brief intermittent walking bouts and random arm motions or static arm status while walking (no clear correlation of motion pattern between arm and leg). 
This paper proposes a low-power step-counting solution utilizing the body area electric field acquired by a novel electrostatic sensing unit, consuming only 87.3 µW of power, hoping to strengthen the robustness of current dominant solution. 
We designed two wearable devices for on-the-wrist and in-the-ear deployment and collected body-area electric field-derived motion signals from ten volunteers. Four walking scenarios are considered: in the parking lot/shopping center with/without pushing the shopping trolley. The step-counting accuracy from the prototypes shows better accuracy than the commercial wrist-worn devices (e.g.,96\% of the wrist- and ear-worn prototype vs. 66\% of the Fitbit when walking in the shopping center while pushing a shopping trolley). We finally discussed the potential and limitations of sensing body-area electric fields for wrist-worn and ear-worn step-counting and beyond. 

\end{abstract}




\maketitle

\section{Introduction}

\thispagestyle{firstpage}

\begin{table*}[htbp]
\centering
\caption{Sensing modalities explored for step counting}
\label{on_device}
\begin{tabular}{ p{0.7cm} p{1.6cm} p{1.2cm}  p{2.1cm} p{2.6cm}  p{2.3cm}   p{4.8cm}}
\hline
\textbf{Works} & \textbf{Sensing Modality}  & \textbf{Deploy- ment}  & \textbf{Algorithm} & \textbf{Accuracy}  &  \textbf{Advantage}  &  \textbf{Limitation}  \\
\hline
\cite{moore2020toward} &  accelerometer & wrist, waist, thigh  & filter, peak detector &  7-11\%(wrist) & compact, low-power, robust  & imperfect in accuracy(wrist) \\
\hline
\cite{ozcan2015robust} &  phone camera & hold in hand  &  image edge detection &  Avg. 96.94\% &  more accurate and robust & lack of practicality caused by privacy issue, power consumption, etc. \\
\hline
\cite{ngueleu2019design} &  pressure & insole  &  cumulative sum of FSR &  95.5-98.5\%(indoor), 96.5-98.0\%(outdoor) &  high accuracy & cost and wearability \\
\hline
\cite{las2021method} &  magnetic & wrist  & extrema detector, comparator  &  98\% &  high accuracy, low cost, compact & arm swing dependant(like accelerometer), environment dependant \\
\hline
\cite{xu2018wistep} &  WIFI & environ- ment  & CSI time frequency analysis &  90.2\%(laboratory), 87.59\%(classroom) &  device-free & environment dependant \\
\hline
\textbf{This works} & Body-area electric field &  wrist/head & filter, peak detector&  Avg. 93\%(wrist), 94\%(head) & deployment-place free, compact, low-power & environment dependant \\
\hline
\end{tabular}
\end{table*}

Daily steps have been proven to have an inverse dose-response relationship with important health outcomes like all-cause mortality and cardiovascular events \cite{kraus2019daily, jayedi2022daily}, thus becoming an important factor in promoting a healthy daily routine\cite{hallam2018happy}. The functionality of step counting has been widely implemented by wearable devices, such as fitness trackers and smartwatches, to provide users with timely feedback on their steps. Accurate step counting is crucial for individuals who rely on wearable devices to quantify their exercise and achieve fitness goals. However, existing step-counting methods based solely on inertial measurement units can be non-robust as they primarily detect the swinging motion of the device attached to a certain body part, such as the wrist, rather than directly measuring the leg lift and drop action. This reliance on the swinging motion of the device can lead to false positive errors, especially in scenarios where wrist movement is not directly correlated with leg movement, e.g., when users wave their arms without doing a step action. A second error source of the current inertial sensor-based approach is a result of brief intermittent walking bouts, e.g. when users walk slowly or when they stop and start walking again, the inertial unit-based counting will be unreliable as the wrist movement is irregular during such period. In \cite{toth2019effects}, the authors evaluated the effects of brief intermittent walking bouts on the step-counting accuracy of wearable devices with commercial trackers worn on the wrist (four), hip (four), and ankle (two). They concluded that most methods required stepping bouts of >6–10 consecutive steps to record steps. Rest intervals of 1–2 seconds were sufficient to break up walking bouts in many methods. The requirement for several consecutive steps in some methods causes an underestimation of steps (false negative) in brief, intermittent bouts.

Some specific studies have been published to evaluate how valid current wearable devices are for measuring steps. In \cite{an2017valid}, the authors used ten commercial wrist-worn activity trackers and collected walking data from thirty-five healthy individuals under three conditions: walking/jogging on a treadmill, walking over-ground on an indoor track, and a 24-hour free-living condition. The result shows that the mean absolute percentage error (MAPE) score for all devices and speeds on a treadmill was 8.2\% against manually counted steps, and the MAPE value was higher for over-ground walking (9.9\%) and even higher for the 24-hour free-living period (18.48\%) on step counts. A similar evaluation was carried out in \cite{tophoj2018validity}, where four commercial activity trackers, Fitbit Surge (FS), Fitbit Charge HR (FC), Microsoft Band 2 (MB), and AD 101NFC Activity Monitor (AD), were evaluated at 2, 4, 4.5, and 5.5km/h. The experiment of twenty healthy participants walked on a treadmill in two trials of 100 steps each showed that the MAPE levels were between 8\% and 6\% for FS, 15\% and 0\% for MB, 7\% and 21\% for FC, and 53\% and 1\% for AD. The biggest inaccuracies were seen at 2km\/h, where AD was underestimated by 53\%. MB, FS, and AD accurately counted steps when participants walked with velocities corresponding to a brisk walk. Walking at lower speeds was not counted accurately. 

To achieve more robust step counting with the wearables and address the gap between inertial sensor readings and real steps, extra step-sensing methods need to be implemented to supply alternative information on steps when the inertial sensor reading fails to capture a real step or distinguish between false positives and true positives.

\section{Related Work}
Robustness study for well-being has been impressively growing in recent years, especially along with the advanced machine learning models \cite{meegahapola2024m3bat}, e.g., by personalizing the models \cite{meegahapola2023generalization}. In this paper, we tried to strengthen the robustness of the current IMU-based step-counting method by introducing an effective complementary sensing approach.    
Several studies using different sensing approaches other than the dominant inertial sensors for step counting have been explored in the past decade, as Table \ref{on_device} lists. In \cite{ozcan2015robust}, the authors present a robust and reliable method for counting footsteps using videos captured with a Samsung Galaxy S4 smartphone. With a comparative study, the proposed method achieved an impressive step counting performance of 3.064\% in average error. 
In \cite{ngueleu2019design}, an insole equipped with pressure sensors was introduced for more accurate step detection. With twelve healthy participants at self-selected and maximal walking speeds in indoor and outdoor settings, the system accurately detected steps with success rates ranging from 95.5\% to 98.5\% (indoor) and from 96.5\% to 98.0\% (outdoor) for self-selected walking speeds.
Despite high accuracy, the implementation cost of such a sensing system limits its usage scenarios and makes it fall behind compared to the current inertial sensor-based solution. A cost-comparative method was also proposed by Patryk et al. in \cite{las2021method}, where a magnetic field sensor-based approach to detect and count steps was described. 
Outdoor experiments showed that the proposed detection method achieves up to 98\% accuracy in step counting. Compared with the inertial sensor solution, this magnetic field-based solution also depends on the arm swing during walking; thus, errors from the brief intermittent walking bouts still exist. 
In \cite{xu2018wistep}, the authors leveraged ubiquitous WiFi signals and proposed a device-free step-counting system named WiStep,
realizing an overall accuracy of 90.2\% in the laboratory and 87.59\% in the classroom. The presented WiStep enjoys the advantage of device-free sensing but is limited by the usage space where efficient CSI information must be extracted from available wifi signals.

In this work, we proposed a novel sensing modality based on the body surface differential electrostatic charge for accurate step counting, trying to address errors of the inertial sensor-based solution, like the false positive during leg-static activities or false negative during intermittent walking. Electrostatic charge on the human body, a fundamental principle of physics, has found intriguing applications for wearables \cite{dheman2022cardiac, irrera2024multisensor}. When a person moves, especially during physical activities, the body generates electrostatic charges due to friction between surfaces. These charges can be harnessed and measured by specialized sensors \cite{bian2024body}. As the body is a perfect conductive object, the surface charge flow can be observed anywhere on the body, enabling cross-body part sensing. Some previous studies have presented some preliminary studies using body-area electric field for human activity recognition \cite{hu2022wearable, bian2019wrist}, but their explorations were limited to indoor environments and lacked a dedicated and comprehensive evaluation in step counting.
Leveraging the principles of electrostatic charge promises to revolutionize the capabilities of wearable devices, ushering in a new era of personalized and accurate activity tracking.

The paper showcases how cross-body part sensing functionality revolutionizes step sensing by eliminating the need for specific deployment positioning, thus offering a novel method for strengthening robustness and achieving advanced accuracy in step counting. Overall, we bring the following contributions in this work:

\begin{enumerate}
\item We designed the body-area electric field sensing module based on an electrostatic sensor, being able to sense the static electric field variation during body motion. The module is then integrated into two wrist-worn and ear-worn prototypes for body motion-sensing exploration.
\item We collected the body-area electric field variation data set from ten volunteers walking in/outside a shopping center with/without the shopping trolley and evaluated the step-counting performance using the wrist-worn and ear-worn prototypes. Results show that the static body-area electric field sensing modality supplies an averaged accuracy of 93\%(wrist-worn) and 94\%(head/ear-worn) step counting, outperforming two commercial products, which infers the feasibility of electric field-based step counting. We further discussed the limitations of the current work and proposed future work in sensing fusion.
\end{enumerate}

\begin{figure}[hbt]
\graphicspath{{./Figures/}}
\centering
\includegraphics[width=0.99\linewidth, height = 4cm]{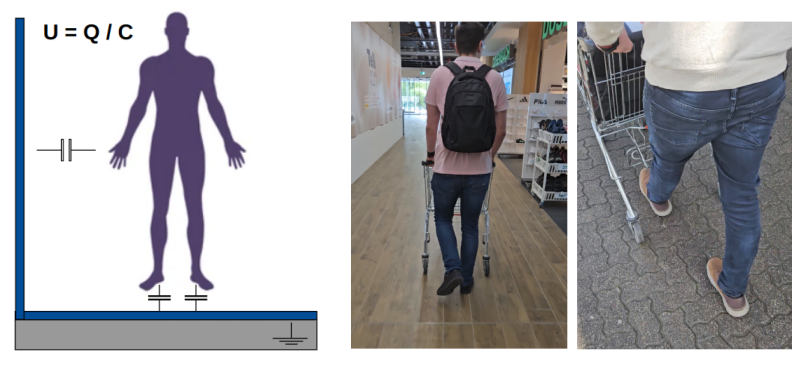}
\caption{Left: Capacitor-like modeling of the human body describes the Body-area electric field, capacitive coupling through the air, shoe soles, and ground. $U$ = human body surface potential, $Q$ = human body surface charge, $C$ = human body capacitance. Right: Data collection in/out of a shopping center while pushing a shopping trolley}
\label{Model}
\end{figure}

\section{Methods}

\subsection{Principle}

Body-area electric field, on a narrow scale, infers the passive static electric field between the human body and surroundings (especially the ground), also defined as human body capacitance \cite{jonassen1998human}. Given the fact that approximately sixty percent of the human body comprises water, it inherently exhibits conductive properties, being able to store charges. Being insulated by the clothing (especially the shoes), a surface potential related to the environment arises as a result of the charge distribution difference between the body and the environment, thus forming a static, passive body-area electric field \cite{jonassen1998human, using2022bian}. Fig. \ref{Model} (left) depicts a capacitor-like model of the human body, where $U$ = human body surface potential, $Q$ = human body surface charge, and $C$ = human body capacitance. During motion actions, like steps, the relative distance/overlapping area change (between the body and surrounding) will cause the variation of the human body capacitance, which results in the on-body surface charge flow, also represented as body-area electric field variation. By sensing the body charge flow pattern, certain body activities could be inferred. Recent years have shown an increasing trend in such passive body-area static electric field sensing exploration, like random motion sensing \cite{cohn2012ultra}, indoor positioning \cite{tang2019indoor}, exercise recognition \cite{bian2019passive, bian2022exploring}, etc., based on the fact that such sensing is ultra-low power, low cost, noninvasive, and enjoying the advantage of deployment position-free compared to traditional inertial unit, being a promising approach for wearable devices to accomplish certain tasks in human activity recognition and human-computer interaction.

\begin{figure}[hbt]
\graphicspath{{./Figures/}}
\centering
\includegraphics[width=0.95\linewidth, height = 8.0cm]{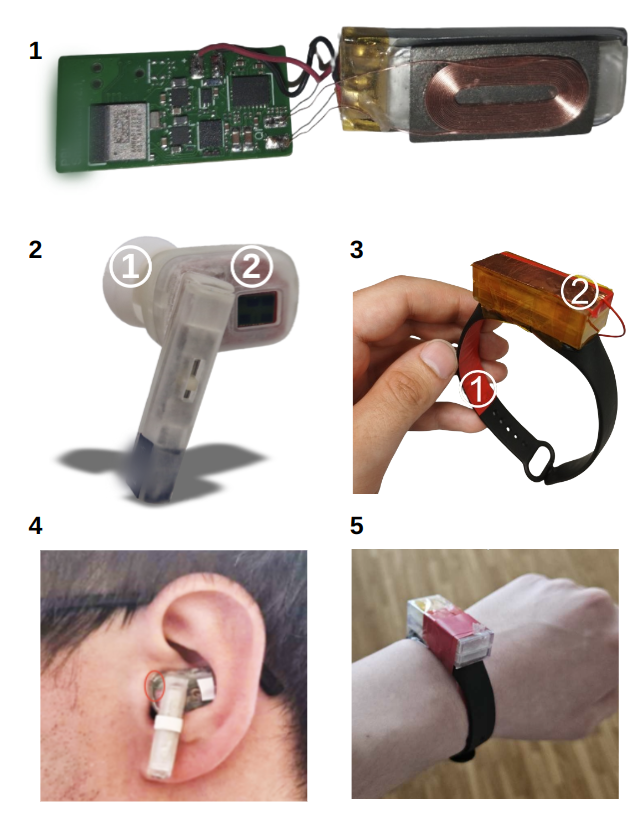}
\caption{Hardware implementation. 1, The main board, including the key components of a Qvar sensing unit, a low-power MCU, a lithium battery, and a wireless charging module. 2, In-ear deployment where the main board is in the case, and the electrode is underneath the cap. 3, Wrist-worn deployment where the main board is in the case and the electrode is underneath the band. 4/5, the real device-wearing scenarios.}
\label{hardware}
\end{figure}


\begin{figure*}[htb]
\graphicspath{{./Figures/}}
\centering
\includegraphics[width=0.9\linewidth, height = 6.0cm]{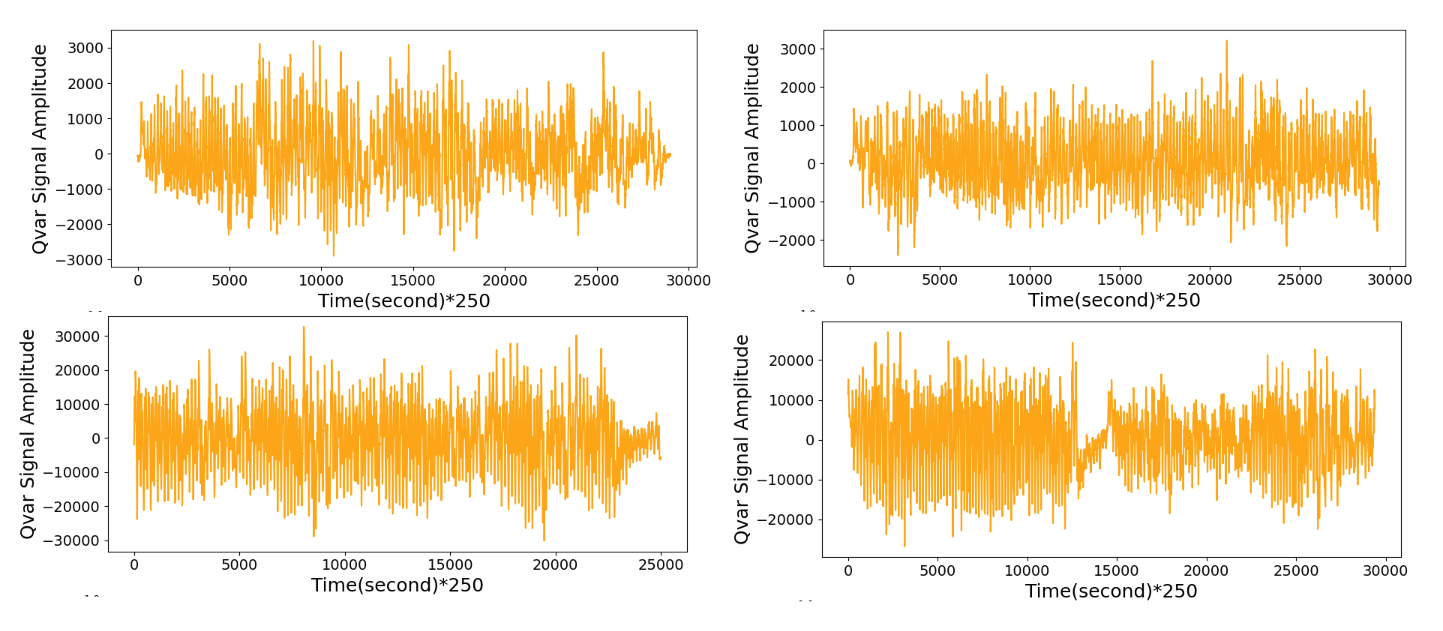}
\caption{Raw signal in the four sub-sessions from a volunteer's single whole session. Top left: at the parking lot without pushing the shopping trolley; top right: at the parking lot while pushing the shopping trolley; bottom left: in the shopping center while pushing the shopping trolley: bottom right: in the shopping center without pushing the shopping trolley. The unit of Y-axis is 1.8/2$^{16}$(volt)}
\label{Raw_signal}
\end{figure*}

\subsection{Hardware Implementation}

Fig. \ref{hardware} shows the hardware implemented in this study. The key components on the main board include a Qvar sensing unit, a low-power u-blox ANNA-B112 module, a lithium battery, and a wireless charging module. Qvar \cite{stm_manual}, stands for electric charge (= Q) variation (= var), a newly released electrostatic sensor from STMicroelectronics, has an electrical potential sensing channel able to measure the quasi-electrostatic potential changes, enabling applications such as contact and no-contact human motion detection, human presence sensing, and user interface. Researchers have proved its effectiveness in hand gesture recognition \cite{reinschmidt2022realtime} with a wrist-worn band composing multiple Qvar units and inertial units in a sensor fusion way, and cardiac monitoring \cite{dheman2022cardiac} incorporated into a wearable chest strap that can be integrated seamlessly under clothes. Qvar is an ultra-low power unit, consuming only 87.3 µW of power and lasting multiple weeks of operation using a coin cell battery. The processing unit on the board, ANNA-B112, is a System-in-package (SiP) module around the ultra-low power and high-performance nRF52832 System-on-Chip (SoC), featuring a state-of-the-art BLE 5.4 interface, as well as an ARM Cortex-M4F MCU. A similar but miniaturized design of the main board is designed to fit in a 3D-printed ear-worn case, where the sensing electrode is put underneath the cap of the case, forming a contactless in-the-ear sensing prototype, as 3/4 in Fig. \ref{hardware} show. Likewise, a wrist-worn prototype is designed with the sensing electrode underneath the band and covered by insulating tape so that the wrist-worn prototype senses the body surface potential variation also in a contactless way. The Qvar signal data was sampled at 240 Hz and transmitted wirelessly through the BLE module to a nearby personal computer.

\subsection{Data Collection}

To verify the capability of step counting with the developed prototypes in the wild, we chose a general scenario of visiting a shopping center as a practical everyday scenario. Ten young adults (five females and five males, ages 21 to 32) participated in the data collection activity, and each of them performed one session of data collection. During the data collection, the subject wore our prototypes on the wrist and in the ear and two additional commercial activity trackers (Fitbit Inspire 3 and Xiaomi Smart Band 8, both the latest in their series) on the other wrist. This setup is simply for user convenience and will not influence the general counting result, as each submission lasts around two minutes, and random left and right swing differences can be ignored. Each session was composed of four sub-sessions. Starting from the parking lot outside the shopping center, the subject walked normally for around two minutes. Then, a shopping trolley was picked up and used in the shopping center; this period also lasted around two minutes. The third sub-session was normal walking without the shopping trolley in the shopping center. Finally, the subject had another around two minutes of normal walking outside the shopping center in the parking lot. Such a configuration aims to test the sensor's ability to count steps when, first, the wrist swings normally during walking and relatively static during pushing a shopping trolley; second, the environment changes, as this electric field-based sensing modality describes the static electric relations between human body and environment, thus, the environment plays a role in signal quality. During each submission, a second accompanying subject used a smartphone to record the leg actions, aiming to provide the ground truth. At the start and end of each sub-session, the step readings of the commercial products were recorded for comparison. Fig. \ref{Model}(right) are the screenshots of the recorded data collection video, showing the inside/outside normal walking while pushing a shopping trolley. In the parking lot, the ground is composed of concrete bricks, and in the shopping center, it is composed of wood. Fig. \ref{Raw_signal} depicts one session of the Qvar signal composed of the four sub-sessions. As can be seen, the amplitude of the sensed surface potential variation in the parking lot and shopping center differs a lot (as the body-area electric field is an interactive signal between the body and environment), but the step waves remain, and we extract the step numbers from those wave signals.


\begin{figure}[hbt]
\graphicspath{{./Figures/}}
\centering
\includegraphics[width=0.99\linewidth, height = 9.0cm]{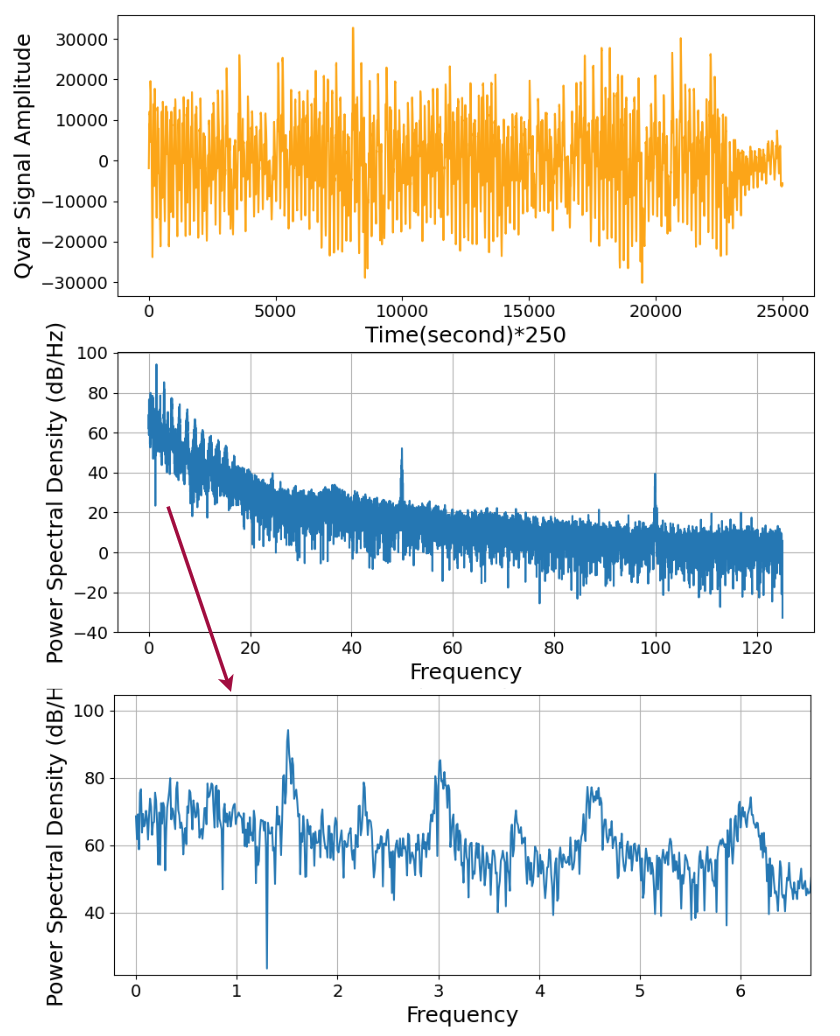}
\caption{One sub-session of Qvar signal and the corresponding power spectral density describing the distribution of power into frequency components, which indicates the dominant frequency during walking}
\label{PSD}
\end{figure}


\section{Evaluation}

\begin{figure}[]
\graphicspath{{./Figures/}}
\centering
\includegraphics[width=0.99\linewidth, height = 9.0cm]{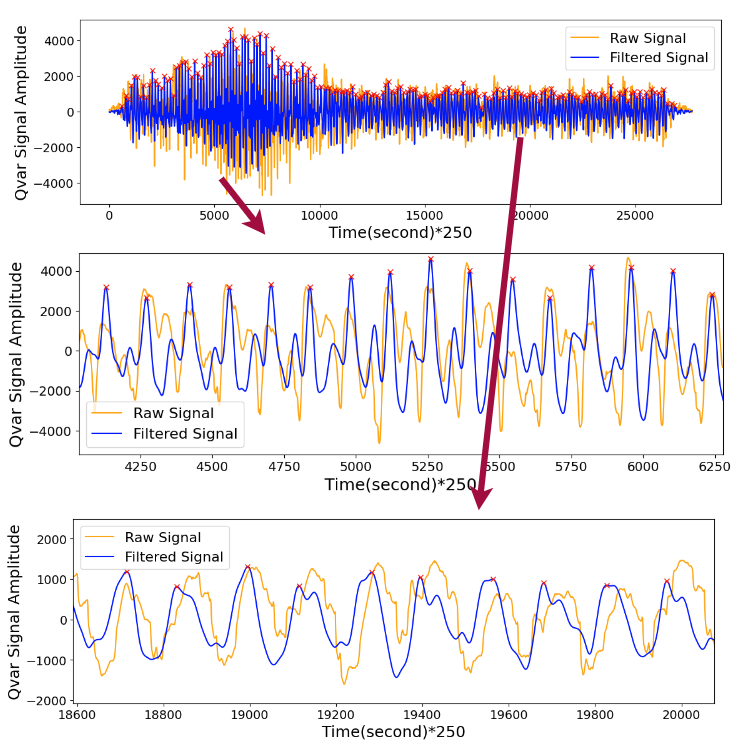}
\caption{Filtered signal marked with the detected peaks}
\label{Filtered_signal}
\end{figure}

Like the commonly used algorithms dealing with inertial sensor data for step counting \cite{khedr2017smartphone, yang2015accurate, brajdic2013walk}, here we use a bandpass filter and a peak detection algorithm for step extraction. A lot of studies have concluded that the typical walking frequency is below 2.0 Hz during a normal walk \cite{yang2015accurate, heinemann2017damping, nguyen2011evaluation}. For example, in \cite{nguyen2011evaluation}, the authors described a biomedical research program that used the GAITRite electronic walkway system with pressure-activated sensors to investigate the basic spatial and temporal gait measures. The participants completed a series of walks at self-selected free (normal), fast and slow gait speeds across the GAITRite walkway. The obtained mean step frequency for the normal walk is 1.98 Hz with a standard deviation of 0.13 Hz, which compares well with values reported by Bachmann and Ammann(1987) \cite{bachmann1987vibrations} and the European Commission(2006) \cite{null}. 
The frequency feature of our collected step signal based on the body-area electric field also matches previous findings. Fig \ref{PSD} shows the power spectral density (PSD) of a one-subsession Qvar signal, depicting the distribution of power with frequency components and indicating the dominant frequency during normal walking. We set the Butterworth bandpass filter with a lower cutoff frequency of 0.5Hz, a higher cutoff frequency of 2.5Hz, and the order with 5. Following the filter, we applied the peak detection algorithm, which is implemented by using the $find\_peaks$ in the scipy package \cite{2020SciPy-NMeth}, which has multiple parameters, and we only use the prominence (vertical difference between the peaks height itself and its lowest contour line) and distance (required minimal horizontal distance in samples between neighboring peaks). The tested prominence ranges from 200 to 500 with a step of 50, and the tested distance ranges from 50 to 200 with a step of 50. A grid searching method is used for each session to find the best parameter set and finally choose the best set with major voting. Such a subject-undependable fine-tuning will maximally guarantee the algorithm's fairness, which has been a growing concern in recent well-being studies \cite{yfantidou2023beyond, meegahapola2024m3bat, meegahapola2023generalization}.
Such a straightforward data mining pipeline will lead to a window-based real-time step counting. Fig \ref{Filtered_signal} shows the depicted steps with red stars along with the filtered and raw signal in a parking lot sub-session. The amplitude variation during the sub-session indicates the surrounding instability from the environment. Since the step action (distance between feet and ground) strongly influences the capacitance in between, signals of steps still dominate the changes in sensor reading.

Table \ref{accuracy} listed the step counting result of each sub-session, including the result of the two commercial devices (Fitbit Inspire 3 and Xiaomi Smart Band 8). The ear-worn Qvar unit from the first, seventh, and eighth subject in the first submission (normal walking in the parking lot without pushing the shopping trolley) is missed because of an unexpected break in the device-computer BLE connection. The accuracy is calculated by the following equation:

\begin{equation}
    Accuracy = 1 - \frac{abs(Measurement - Truth)}{Truth}
\end{equation}

Finally, the accuracies of all subjects are averaged as the final result of each prototype and product. As can be seen, the Qvar-based prototype outperforms in the scenarios when the subject is pushing the shopping trolley, especially in the shopping center, with an average accuracy of 96\% from both the wrist-worn and ear-worn prototypes. The Xiaomi Band basically lost the step counting accuracy when subjects are pushing the shopping trolley, this is probably because that their step counting algorithm neglect the tiny regular waves when the wrist is in a near-static status. However, it shows the best-averaged accuracy during normal walking without the shopping trolley. The wrist-worn prototype gives better counting accuracy in the shopping center, while the ear-worn prototype performs better when pushing the shopping trolley, but overall, they always win the Fitbit with all accuracies above 90\%, which proves, first, the deployment location-free character (cross-body part) of body-area electric field-based motion sensing, as the body is a complete object, it doesn't matter where the sensing unit is deployed on body; The close accuracies between the two prototypes also support this principle. Second, the body-area electric field-based sensing can be used for accurate step counting in certain scenarios, especially when the inertial unit-based solution fails, like during a very slight wrist swing.

\begin{table}[t!]
\centering
\caption{Step counting result from commercial devices and introduced module deployed on wrist and head (step number / absolute accuracy)}
\label{accuracy}
\begin{tabular}{ p{0.8cm} p{0.8cm}  p{1.2cm} p{1.2cm} p{1.2cm} p{1.2cm}  }
\hline
\textbf{Subject} & \textbf{Truth} & \textbf{Xiaomi} & \textbf{Fitbit} & \textbf{WristQvar} & \textbf{EarQvar} \\
\hline
\multicolumn{6}{c}{\textbf{In parking lot without shopping trolley}}\\\hline
1 & 160 & 169/0.94 & 164/0.98  & 160/1.00 & xx/xx \\\hline
2 & 160 & 164/0.98 & 174/0.91  & 170/0.94 & 165/0.97 \\\hline
3 & 159 & 152/0.96 & 180/0.87  & 177/0.89 & 115/0.72 \\\hline
4 & 164 & 167/0.98 & 179/0.91  & 171/0.96 & 160/0.98 \\\hline
5 & 159 & 160/0.99 & 181/0.86  & 169/0.94 & 159/1.00 \\\hline
6 & 215 & 211/0.98 & 247/0.85  & 160/0.74 & 176/0.82 \\\hline
7 & 204 & 206/0.99 & 233/0.86  & 210/0.97 & xx/xx \\\hline
8 & 205 & 228/0.89 & 226/0.90  & 150/0.73 & xx/xx \\\hline
9 & 176 & 173/0.98 & 195/0.89  & 178/0.99 & 177/0.99 \\\hline
10 & 166 & 166/1.00 & 184/0.89  & 168/0.99 & 165/0.99 \\\hline
\textbf{Avg.} &  & \textbf{0.97} &  \textbf{0.89} & \textbf{0.91} & \textbf{0.92} \\\hline
\multicolumn{6}{c}{\textbf{In parking lot with shopping trolley}}\\\hline
1 & 165 & 83/0.50 & 166/0.99  & 168/0.98 & 167/0.99 \\\hline
2 & 167 & 23/0.14 & 200/0.80  & 183/0.90 & 177/0.94 \\\hline
3 & 154 & 0/0.00 & 131/0.85  & 176/0.86 & 159/0.97 \\\hline
4 & 158 & 9/0.06 & 135/0.85  & 163/0.97 & 157/0.99 \\\hline
5 & 167 & 0/0.00 & 170/0.98  & 175/0.95 & 171/0.98 \\\hline
6 & 194 & 64/0.33 & 213/0.90  & 156/0.80 & 161/0.83 \\\hline
7 & 202 & 71/0.35 & 234/0.84  & 167/0.83 & 181/0.90 \\\hline
8 & 198 & 77/0.39 & 224/0.87  & 193/0.97 & 170/0.86 \\\hline
9 & 165 & 30/0.18 & 197/0.81  & 171/0.96 & 169/0.98 \\\hline
10 & 167 & 11/007 & 183/0.90  & 158/0.95 & 173/0.96 \\\hline
\textbf{Avg.} &  & \textbf{0.20} &  \textbf{0.88} & \textbf{0.92} & \textbf{0.94} \\\hline
\multicolumn{6}{c}{\textbf{In shopping center with shopping trolley}}\\\hline
\textbf{Avg.} &  & \textbf{0.01} &  \textbf{0.66} & \textbf{0.96} & \textbf{0.96} \\\hline
\multicolumn{6}{c}{\textbf{In shopping center without shopping trolley}}\\\hline
\textbf{Avg.} &  & \textbf{0.95} &  \textbf{0.83} & \textbf{0.94} & \textbf{0.93} \\\hline


\end{tabular}
\end{table}

\section{Discussion}

This work adopted a shopping activity composed of four subsections to explore the potential of body-area electric field-based step counting and demonstrate its feasibility. Nevertheless, it must be acknowledged that this sensing modality is contingent upon the environment, as evidenced by the amplitude disparities observed inside and outside the shopping center. The presence of numerous electric appliances within the center amplifies the variations in the electric field, rendering them more pronounced. We have observed that in some places, for instance, in the huge empty training yard, the explored signal is too weak to lose the motion sensing ability. This indicates that such a sensing module might lead to issues in robustness. Thus, future work will focus on explorations towards the robustness by, first, designing a front end for Qvar or a new front end for electric field variation sensing so that the sensing unit has a larger input impedance to sense very slight electric field variation signal and increase the signal-to-noise ratio at the meantime; second, developing a fusion solution combining the inertial unit and body-area electric field sensing unit, as they are both ultra-low power, low-cost ideal wearable components and shares different advantages in motion sensing, a fusion method will potentially supply the best capability of human activity recognition like step counting. 

\section{Conclusion}

This work proposed the body-area electric field-based sensing for step counting. The main goal of the paper is to evaluate this emerging sensing feature to improve step-counting accuracy in wearable devices.  Two prototypes (a wrist-worn and an ear-worn) were designed to evaluate the step-counting proposal and verify the cross-body part motion sensing technique. Experimental results have demonstrated counting accuracies of 91\% to 96\%  in the shopping activity composed of four subsessions(inside/outside a shopping center and with/without pushing a shopping trolley), which outperforms the Fitbit Inspire 3 in all four subsessions and Xiaomi Smart Band 8 in two submissions. The counting result indicates the feasibility of using body-area electric field sensing for accurate step counting, considering the false positive and false negative errors of traditional inertial sensor-based, as well as other activity sensing. Future work will focus on the sensing front-end improvement and sensor fusion way to increase the robustness of this sensing modality. All materials from this work will be open-sourced aiming to promote further exploration in body-area electric field-based motion sensing.

\begin{acks}
This work was supported by the CHIST-ERA project ReHab(20CH21-203783).
\end{acks}

\bibliographystyle{ACM-Reference-Format}
\bibliography{sample-base}


\end{document}